\documentstyle[12pt,a4,epsf]{article}

\def\tr{\mathop{\rm tr}\nolimits}

\makeatletter
\@addtoreset{equation}{section}
\makeatother

\newcommand{\VEV}[1]{\left\langle #1 \right\rangle}

\newcommand{\Mp}{M_P}

\begin{document}
\begin{titlepage}

\begin{flushright}
hep-ph/0111205\\
KUNS-1746\\
\today
\end{flushright}

\vspace{4ex}

\begin{center}
{\large \bf
Gauge Coupling Unification  \\
with Anomalous $U(1)_A$ Gauge Symmetry
}

\vspace{6ex}

\renewcommand{\thefootnote}{\alph{footnote}}

Nobuhiro Maekawa\footnote
{E-mail: maekawa@gauge.scphys.kyoto-u.ac.jp
}

\vspace{4ex}
{\it Department of Physics, Kyoto University,\\
     Kyoto 606-8502, Japan}\\
\end{center}

\renewcommand{\thefootnote}{\arabic{footnote}}
\setcounter{footnote}{0}
\vspace{6ex}

\begin{abstract}

Recently we proposed a natural scenario of grand unified 
theories with anomalous $U(1)_A$ gauge symmetry, in which 
doublet-triplet splitting is realized in $SO(10)$ 
unification using Dimopoulos-Wilczek mechanism and realistic 
quark and lepton mass matrices can be obtained in a simple way.
The scenario has an additional remarkable feature that
the symmetry breaking scale and the mass spectrum of super 
heavy particles are determined essentially by anomalous 
$U(1)_A$ charges. Therefore once all the anomalous $U(1)_A$
charges are fixed, the gauge coupling flows 
can be calculated. We examine several models in which the 
gauge coupling unification is realized. 
Examining the conditions for the coupling
unification, we show that when all the 
fields except those of the minimal SUSY standard model become 
super-heavy, the unification scale generically becomes 
just below the usual GUT scale $\Lambda_G\sim 2\times 10^{16}$ 
GeV and the cutoff scale becomes around $\Lambda_G$. 
Since the lower GUT scale leads to shorter life time of nucleon, 
the proton decay via dimension six operator $p\rightarrow e^+\pi^0$
 can be seen in future experiment.
On the other hand, the lower cutoff scale than the Planck scale may 
imply the existence of extra dimension in which only gravity modes 
can propagate.

\end{abstract}

\end{titlepage}


\section{Introduction}
There is strong evidence supporting grand unified theories 
(GUT)\cite{georgi}, 
in which the quarks and leptons are unified
in several multiplets in a simple gauge group.
They explain various matters that cannot be understood within the
standard model: the miracle of anomaly cancellation between quarks
and leptons, the hierarchy
of gauge couplings, charge quantization, etc.
The three gauge groups in the
standard model are unified into a simple gauge group at a GUT scale, 
which is considered to be just below the Planck scale.
On the other hand, the GUT scale destabilizes the weak scale. 
One of the most promising
ways to avoid this problem is to introduce supersymmetry (SUSY).
One of the most important successes of SUSY is regarded as
the gauge coupling unification. In the minimal SUSY standard
model (MSSM), three gauge couplings meet at a single scale 
$\Lambda_G\sim 2\times 10^{16}$ GeV. 

However, it is not easy to obtain a realistic SUSY 
GUT.\cite{SUSYGUT}
First, it is difficult to obtain realistic
fermion mass matrices in a simple way.
In particular, unification of quarks and leptons puts strong
constraints
on the Yukawa couplings.
But concerning the fermion masses,
recent progress in neutrino experiments\cite{SK} 
provides important information on family structure.
There are several impressing 
works\cite{Sato,Nomura,IKNY,Bando,Barr,Shafi} in which 
the large neutrino mixing angle is realized within GUT framework.
It is now natural to examine
$SO(10)$ and higher gauge groups, because they allow for 
every quark and lepton,
including the 
right-handed neutrino, to be unified in a single multiplet, which is
important in addressing neutrino masses.

Second, one of the most difficult obstacles is
the ``doublet-triplet(DT) splitting problem".
Generally, a fine-tuning is required to obtain the light 
$SU(2)_L$ doublet Higgs multiplet of the weak scale while 
keeping the triplet Higgs sufficiently heavy to suppress 
the dangerous proton decay. 
There have been several attempts to solve this  
problem.\cite{DTsplitting,DW} 
Among them, the Dimopoulos-Wilczek mechanism is
a promising way to realize
DT splitting in the $SO(10)$ SUSY 
GUT.\cite{DW,BarrRaby,Chako,complicate}

Finally, there is a rather theoretical problem, which has not
been emphasized so much in the literature. If we adopt an ajoint
Higgs field $A$ to break the GUT gauge group, the superpotential
is generically given by $W=\sum_n^\infty A^n$. In the vacua, 
the GUT gauge group is generically broken to $U(1)^r$, where
$r$ is the rank of the GUT gauge group. It is unnatural
to obtain the standard gauge group 
$SU(3)_C\times SU(2)_L\times U(1)_Y$ by this superpotential. 
At least for $SU(5)$ unification, we can impose renormalizability 
to avoid this problem. 
Then the superpotential becomes $W=A^2+A^3$, which naturally
gives the standard gauge group below the GUT scale.
However, for $SO(10)$ or $E_6$ unification, it is not workable because
$A^3$ is not allowed under the gauge symmetry. 
Moreover, in the context of Wilsonian renormalization group,
renormalizability is not a principle to be imposed, but a
resulting feature which low energy effective theories happen to
obtain. When the cutoff scale is much higher than the 
typical scale of the theory, higher dimensional operators become 
irrelevant, and only a few operators are important to determine 
the low energy physics. However, since the 
GUT scale is near the Planck scale, it is not natural to impose 
the renormalizability on the grand unified theories.
Therefore we have to answer the question why the non-Abelian gauge
group remains at the low energy scale, or, why the allowed
interactions in the superpotential are restricted.

Recently we proposed
 a scenario of $SO(10)$ (also $E_6$) grand unified theory (GUT) 
 with anomalous $U(1)_A$ gauge symmetry\cite{maekawa,BM},
 which has the following interesting features;
\begin{enumerate}
\item
The interaction
is generic, namely, all the interactions, which are allowed 
by the symmetry, are introduced. Therefore, once we 
fix the field contents with their quantum numbers (integer), 
all the interactions are determined except the coefficients
of order one.
\item
Even with generic interaction, non-Abelian gauge group at the
low energy scale can be obtained. (Necessary restriction to the
superpotential is realized by SUSY zero mechanism.)
\item
The Dimopoulos-Wilczek mechanism, which realizes the 
doublet-triplet (DT) splitting,  is naturally embedded.
\item
The proton decay via the dimension-five operator is suppressed.
\item
Realistic quark and lepton mass matrices can be obtained in a simple
way. In particular, in the neutrino sector, bi-large neutrino mixing is 
realized.
\item
The symmetry breaking scales are determined by the anomalous $U(1)_A$
charges.
\item
The mass spectrum of the super heavy particles is fixed by the anomalous
$U(1)_A$ charges.
\end{enumerate}
As a consequence of the above features, the fact that the GUT scale is
smaller than the Planck scale leads to modification of the undesired
GUT relation between the Yukawa couplings 
$y_\mu=y_s$ (and also $y_e=y_d$)
while preserving $y_\tau=y_b$. 

The anomalous $U(1)_A$ gauge symmetry,\cite{U(1)}
whose anomaly is cancelled by the Green-Schwarz 
mechanism,\cite{GS}
 plays an essential role
in explaining the DT splitting mechanism at the unification
scale and the restriction of the interactions in the superpotential
as well as in reproducing Yukawa 
hierarchies\cite{Ibanez,Ramond,Dreiner}.
Unfortunately to solve the DT splitting problem, several
super-heavy particles become lighter than the GUT scale. Generically, 
the existence of these fields destroys the success of the 
gauge coupling unification.
However, the spectrum of the super heavy particles are determined
by the anomalous $U(1)_A$ charges, we can calculate the running 
gauge couplings and easily examine whether these couplings meet
at an unified scale or not. In other words, requirement of gauge 
coupling
unification gives some constraints on the anomalous $U(1)_A$ charges.
In this paper, we examine the constraints and try to find out models
in which gauge coupling unification is realized. It is suggestive that
when all the other fields but those of the MSSM become super-heavy,
only a condition leads to the gauge coupling unification. It is 
interesting
that the cutoff scale becomes the usual GUT scale $\Lambda_G$ 
and the unified scale becomes just below the scale $\Lambda_G$. 

In section 2, we explain how the SUSY vacua are determined in the 
anomalous $U(1)_A$ framework. Using this argument, we recall
the discussion of 
the DT splitting mechanism in section 3, and the resulting mass spectrum
of super-heavy particles in section 4. 
In section 5, we review how to determine the anomalous $U(1)_A$ charges
to realize Quark and Lepton mass matrices and bi-large neutrino mixing 
angles. These have been discussed in 
Ref.~\cite{maekawa}. In section 6, we briefly explain how to
solve the $\mu$ problem in our scenario, following the discussion in 
Ref.~\cite{maekawa2}. In section 7, we discuss the conditions for gauge 
coupling unification and in section 8, we examine several models in which
these conditions are satisfied. 

\section{Vacuum determination}
In this section, we explain how the vacua of the Higgs fields are
determined by the anomalous
$U(1)_A$
quantum numbers. 

First, we show that none of the fields with positive anomalous
$U(1)_A$ charge acquire non-zero VEV if the Froggatt-Nielsen (FN) 
mechanism
\cite{FN}
acts effectively in the vacuum.
For simplicity, we here introduce just gauge singlet fields 
$Z_i^\pm$ ($i=1,2,\cdots n_\pm$)
 with charges $z_i^\pm$ ($z_i^+>0$ and $z_i^-<0$).
From the $F$-flatness conditions of the superpotential,
we get $n=n_++n_-$ equations  plus one $D$-flatness condition,
\begin{equation}
 \frac{\delta  W}{\delta Z_i}=0, \qquad D_A=g_A
      \left(\sum_i z_i |Z_i|^2 +\xi^2 \right)=0,
\label{eq:fflat}
\end{equation}
where 
$\xi^2=\frac{g_s^2\tr Q_A}{192\pi^2} (\equiv \lambda^2 \Lambda^2)$
is the coefficient of Fayet-Iliopoulos $D$-term and
$\Lambda$ is a cutoff scale of the theory.
Throughout this paper we use a unit in which $\Lambda=1$ and denote 
all the
superfields with uppercase letters and their anomalous $U(1)_A$ charges
with the corresponding lowercase letters.
At first glance, these look to be over determined. However,
 the $F$-flatness
conditions are not independent, because the gauge invariance of the
superpotential $W$ leads to the relation
\begin{equation}
\frac{\delta  W}{\delta Z_i}z_iZ_i=0.
\label{constraint}
\end{equation}
Therefore, generically a SUSY vacuum with $\VEV{Z_i}\sim \Lambda$ 
exists
(Vacuum a),
because the coefficients of the above conditions are generically
of order 1. 
However, if $n_+\leq n_-$, we can choose another vacuum (Vacuum b)
with $\VEV{Z_i^+}=0$, which automatically satisfies the $F$-flatness
conditions $\frac{\delta  W}{\delta Z_i^-}=0$. Then the $\VEV{Z_i^-}$
are determined by the $F$-flatness conditions
$\frac{\delta  W}{\delta Z_i^+}=0$ with the constraint
(\ref{constraint}) and the $D$-flatness condition $D_A=0$.
Note that if $\xi<1$ (In this paper, we take $\xi=\lambda\sim 0.2$), 
the VEVs of $Z_i^-$ are less than the
cutoff scale. This can lead to the Froggatt-Nielsen mechanism.
If we fix the normalization of $U(1)_A$ gauge symmetry so that
the largest value $z_1^-$ in the negative charges $z_i^-$ equals -1, 
then the VEV of the field $Z_1^-$ is determined from $D_A=0$ as 
$\VEV{Z_1^-}\sim \lambda$,
which breaks $U(1)_A$ gauge symmetry. (We explain later why we take
the field $Z_1^-$ with the largest charges $z_1^-$.) 
In the following, we denote the Froggatt-Nielsen field $Z_1^-$ as 
$\Theta$.
On the other hand,
other VEVs are determined by the $F$-flatness conditions of $Z_i^+$ as
$\VEV{Z_i^-}\sim \lambda^{-z_i^-}$, which is shown below.
Since $\VEV{Z_i^+}=0$, it is sufficient to examine the terms linear
in $Z_i^+$ in the superpotential in order to determine 
$\VEV{Z_i^-}$. Therefore, in general
the superpotential to determine the VEVs can be written
\begin{eqnarray}
W&=&\sum_i^{n_+}W_{Z_i^+},\\
W_{Z_i^+}&=& \lambda^{z_i^+}Z_i^+\left(\sum_j^{n_-}\lambda^{z_j^-}Z_j^-
+\sum_{j,k}^{n_-}\lambda^{z_j^-+z_k^-}Z_j^-Z_k^-+\cdots \right)\\
&=&\sum_i^{n_+} \tilde Z_i^+\left(\sum_j^{n_-}\tilde Z_j^-
+\sum_{j,k}^{n_-}\tilde Z_j^-\tilde Z_k^-+\cdots \right),
\end{eqnarray}
where $\lambda=\VEV{\Theta}$ and $\tilde Z_i\equiv \lambda^{z_i}Z_i$.
The $F$-flatness conditions of the $Z_i^+$ fields require
\begin{equation}
\lambda^{z_i^+}\left(1+\sum_j\tilde Z_j^-+\cdots\right)=0,
\end{equation}
which generally lead to solutions $\tilde Z_j^-\sim O(1)$
if these $F$-flatness conditions determine the VEVs.
Thus the F-flatness condition requires
\begin{equation}
   \VEV{ Z_j^-} \sim O(\lambda ^{-z_j^-}).
\label{VEV}
\end{equation}
Note that if $n_+=n_-$, generically all the VEVs of $Z_i^-$ are
fixed, therefore there appears no flat direction in the potential.
It means that there is no massless field.
On the other hand, if $n_+<n_-$, generally the $n_+$ equations of
$F$-flatness and $D$-flatness conditions do not determine all the
VEVs of $n_-$ fields $Z_i^-$. Therefore, there are flat directions
in the potential, namely there must be some massless fields.
Roughly speaking, if we would like to realize no massless mode
in the Higgs sector, $n_+=n_-$ must be imposed in Higgs
sector.\footnote{
This rough argument of number counting is based on an assumption 
that the Higgs sector has no other structure by which the freedom 
of $F$-flatness conditions is reduced as realized in this section by
taking $\VEV{Z^+}=0$. Such a structure is easily realized by 
imposing some symmetry, for example, $Z_2$ parity or R parity.
However, it is obvious that $n_+\geq n_-$ is required 
to make all the Higgs fields super-heavy.}

Here we have examined the VEVs of singlets fields, but generally
the gauge invariant operator $O$ with negative charge $o$ has
non-vanishing VEV $\VEV{O}\sim \lambda^{-o}$ if the $F$-flatness
conditions determine the VEV. For example, let us introduce spinors
$C({\bf 16})$ and $\bar C({\bf \overline{16}})$ of $SO(10)$. 
The VEV of the gauge singlet operator $\bar CC$ is estimated as
$\VEV{\bar CC}\sim \lambda^{-(c+\bar c)}$. The $D$-flatness condition
of $SO(10)$ gauge theory requires 
\begin{equation}
\VEV{C}=\VEV{\bar C}\sim \lambda^{-(c+\bar c)/2}.
\end{equation}
Note that these VEVs are also determined by the anomalous $U(1)_A$
charges but they are different from the naive expectation
$\VEV{C}\sim\lambda^{-c}$. 
This is because the $D$-flatness condition plays an important
role to fix the VEVs.

Note that if there is another field $Z_i^-$ which has smaller charge
than the FN field $Z_1^-$, then the VEV of $Z_i^-$ becomes
larger than the $\xi$, which is inconsistent with $D$-flatness 
condition. Therefore we have to take the field with the largest negative
charge as the FN field.

If Vacuum a is selected, the anomalous $U(1)_A$ gauge symmetry
is broken at the Planck scale, and the FN mechanism does not act.
Therefore, we cannot know the existence of the $U(1)_A$ gauge symmetry
from the low energy physics. On the other hand, if Vacuum b is
selected, the FN mechanism acts effectively and we can understand
the signature of the $U(1)_A$ gauge symmetry from the low energy
physics. Therefore, it is natural to assume that Vacuum b is
selected in our scenario, in which the $U(1)_A$ gauge symmetry
plays an important role for the FN mechanism. The VEVs of
the fields $Z_i^+$ vanish, which guarantees that the SUSY zero 
mechanism\footnote{
Note that if total charge
of an operator is negative, the $U(1)_A$ invariance forbids 
the operator in the superpotential since the field $\Theta$ with negative 
charge cannot compensate for the negative total charge of the operator 
(SUSY zero mechanism).}
acts effectively.

In summary, 
\begin{enumerate}
\item
Gauge singlet operators with positive total charge have vanishing VEVs,
in order that the FN mechanism acts effectively. This guarantees that 
the SUSY zero mechanism works well.
\item
The $F$-flatness conditions of fields with positive charges determine the
VEVs of singlet operators $O$ with negative total charges $o$ as
$\VEV{O}\sim \lambda^{-o}$, while the $F$-flatness conditions of fields
with negative charges are automatically satisfied.
\item
If the number of the fields with positive charges equals to that of the
fields with negative charges, generically no massless field appears.
\item
General superpotential to determine the VEVs is the following structure
$W=\sum_i W_{Z_i^+}$, where $W_{Z_i^+}$ is linear in the field $Z_i^+$
with positive charges.
\end{enumerate}

\section{Doublet-triplet splitting mechanism}
In this section, we review the mechanism which naturally
realizes the doublet-triplet splitting in  $SO(10)$ unified scenario
\cite{maekawa}.

The contents of the Higgs sector with $SO(10)\times U(1)_A$ gauge 
symmetry 
is given in Table I, where the symbols $\pm$ denote $Z_2$ parity quantum 
numbers.

\vspace{3mm}
\begin{center}
Table I. The typical values of anomalous $U(1)_A$ charges are listed.

\begin{tabular}{|c|c|c|} 
\hline
                  &   Negative charge  & Positive charge \\
\hline 
{\bf 45}          &   $A(a=-2,-)$        & $A'(a'=6,-)$      \\
{\bf 16}          &   $C(c=-4,+)$        & $C'(c'=4,-)$      \\
${\bf \overline{16}}$&$\bar C(\bar c=-1,+)$ & $\bar C'(\bar c'=7,-)$ \\
{\bf 10}          &   $H(h=-6,+)$        & $H'(h'=8,-)$      \\
{\bf 1}           &$Z(z=-3,-)$,$\bar Z(\bar z=-3)$& $S(s=5,+)$ \\
\hline
\end{tabular}

\vspace{5mm}
\end{center}
The adjoint Higgs
field $A$, whose VEV 
$\VEV{A({\bf 45})}_{B-L}=i\tau_2\times {\rm diag}
(v,v,v,0,0)$ breaks $SO(10)$ into
$SU(3)_C\times SU(2)_L\times SU(2)_R\times U(1)_{B-L}$.
 This Dimopoulos-Wilczek form of the VEV plays an
important role in solving the DT splitting problem.
The spinor Higgs fields 
$C$ and $\bar C$, 
that break $SU(2)_R\times U(1)_{B-L}$ into $U(1)_Y$ 
by developing $\VEV{C}(=\VEV{\bar C}=\lambda^{-(c+\bar c)/2}$).
The  Higgs field $H$ contains usual $SU(2)_L$ doublet.
All these Higgs fields must have negative anomalous $U(1)_A$ charges
$a,c,\bar c$ and $h$ to obtain non-vanishing VEVs because only the fields
with negative charge can get non-vanishing VEVs, as discussed in the 
previous section.
On the other hand, in order to give masses to all the Higgs fields,
we have to introduce the fields with positive charges, whose freedom
must be the same as that of the fields with negative charges.\footnote{
Strictly speaking, since some of the Higgs fields are eaten by 
Higgs mechanism, in principle, less number of positive fields can give 
superheavy masses
to all the Higgs fields. Here we do not examine the possibilities.}
Therefore we introduced $A^\prime$, $C^\prime$, $\bar C^\prime$ and
$H^\prime$, which have positive anomalous $U(1)_A$ charges.
Therefore, in a sense, we introduce the minimal Higgs contents here.
It is surprising that the mechanism, in which DT splitting is 
realized, is naturally embedded in such minimal Higgs contents. 

As discussed in the previous section,
since the fields with non-vanishing VEVs have negative charges, only the 
$F$-flatness conditions of fields with positive charge must be taken
into account
for determination of their VEVs. (Generically $c$ or $\bar c$ can be 
positive, since $c+\bar c<0$ is sufficient for non-vanishing VEV.
The following argument does not change significantly if $c$ or $\bar c$
is positive. This is because the terms $C^4$ or $\bar C^4$ does not include
$N_C^4$ or $N_{\bar C}^4$, where $N$ is a neutral component under the
standard gauge group. ) 
We have only to take account of
the terms in the superpotential which contain only one field with 
positive charge. 
Therefore, in general, the superpotential required by determination 
of the
VEVs can be written as
\begin{equation}
W=W_{H^\prime}+ W_{A^\prime} + W_S + W_{C^\prime}+W_{\bar C^\prime}.
\end{equation}
Here $W_X$ denotes the terms linear in the $X$ field, which has 
positive anomalous $U(1)_A$ charge. Note, however, that terms 
including two
fields with positive charge like 
$\lambda^{2h^\prime}H^\prime H^\prime$
give contributions to the mass terms but not to the VEVs.

In the following argument, for simplicity, we neglect
the terms like 
${\bf 16}^4$, ${\bf \overline{16}}^4$, ${\bf 10\cdot 16}^2$,
 ${\bf 10 \cdot \overline{16}}^2$ and ${\bf 1\cdot 10}^2$, 
 even if these terms
are allowed by the symmetry. 
This is because these interactions do not play a significant role
in our argument since they do not include the products of only 
the neutral components under the standard gauge group.
 It is easy to include these terms in
our analysis.

We now discuss the determination of the VEVs.
If $-3a\leq a^\prime < -5a$,
the superpotential $W_{A^\prime}$ is in general
written as
\begin{equation}
W_{A^\prime}=\lambda^{a^\prime+a}\alpha A^\prime A+\lambda^{a^\prime+3a}(
\beta(A^\prime A)_{\bf 1}(A^2)_{\bf 1}
+\gamma(A^\prime A)_{\bf 54}(A^2)_{\bf 54}),
\end{equation}
where the suffixes {\bf 1} and {\bf 54} indicate the representation 
of the composite
operators under the $SO(10)$ gauge symmetry, and $\alpha$, $\beta$ and 
$\gamma$ are parameters of order 1. Here we assume 
$a+a^\prime+c+\bar c<0$
to forbid the term $\bar C A^\prime A C$, which destabilizes the 
DW form of the VEV $\VEV{A}$. 
If we take 
$\VEV{A}=i\tau_2\times {\rm diag}(x_1,x_2,x_3,x_4,x_5)$, the $F$-flatness
of the $A^\prime$ field requires
$x_i(\alpha\lambda^{-2a}+2(\beta-\gamma)(\sum_j x_j^2)+\gamma x_i^2)=0$, 
which gives only two solutions $x_i^2=0$, 
$\frac{\alpha}{(2N-1)\gamma-2N\beta}\lambda^{-2a}$. 
Here $N=1-5$ is the number of $x_i \neq 0$ solutions.
The DW form is obtained when $N=3$.
Note that the higher terms $A^\prime A^{2L+1}$ $(L>1)$ are forbidden by
the SUSY zero mechanism. If they are allowed, 
the number of possible VEVs other than the DW form
becomes larger, and thus it becomes less natural to obtain the DW form. 
This is a
critical point of this mechanism, and the anomalous $U(1)_A$ gauge 
symmetry
plays an essential role to forbid the undesired terms.
It is also interesting that the scale of the VEV is automatically
determined by the anomalous $U(1)_A$ charge of $A$, as noted in the 
previous section.

Next we discuss the $F$-flatness condition of $S$, which determines
the scale of the VEV $\VEV{\bar C C}$. 
$W_S$, which is linear in the $S$ field, is given by
\begin{equation}
W_S=\lambda^{s+c+\bar c}S\left((\bar CC)+\lambda^{-(c+\bar c)}
+\sum_k\lambda^{-(c+\bar c)+2ka}A^{2k}\right)
\end{equation}
if 
$s\geq -(c+\bar c)$.
Then the $F$-flatness condition of $S$ implies $\VEV{\bar CC}\sim 
\lambda^{-(c+\bar c)}$, and the $D$-flatness condition requires 
$|\VEV{C}|=|\VEV{\bar C}|\sim \lambda^{-(c+\bar c)/2}$.
The scale of the VEV is determined only by the charges of 
$C$ and $\bar C$ again.
If we take $c+\bar c=-5$, then we obtain the VEVs of the fields 
$C$ and $\bar C$
as 
$\lambda^{5/2}$, which differ from the expected values $\lambda^{-c}$ 
and
$\lambda^{-\bar c}$ if $c\neq \bar c$.
Note that a composite operator with positive anomalous $U(1)_A$ charge
larger than $-(c+\bar c)-1$ may play the same role as the singlet $S$ if 
such a composite operator exists. (In the above example, there is no 
such composite operator.)

Next, we discuss the $F$-flatness of $C^\prime$ and 
$\bar C^\prime$,
which realizes the alignment of the VEVs $\VEV{C}$ and $\VEV{\bar C}$
and imparts masses on the PNG fields. 
This simple mechanism
was proposed by Barr and Raby.
\cite{BarrRaby}
We can easily assign anomalous $U(1)_A$ charges which allow the 
following superpotential:
\begin{eqnarray}
W_{C^\prime}&=&
       \bar C(\lambda^{\bar c^\prime +c+a}A
       +\lambda^{\bar c^\prime +c+\bar z}\bar Z)C^\prime, \\
W_{\bar C^\prime}&=&
       \bar C^\prime(\lambda^{\bar c^\prime +c+a} A
       +\lambda^{\bar c^\prime +c+z}Z)C.
\end{eqnarray}
The $F$-flatness conditions $F_{C^\prime}=F_{\bar C^\prime}=0$ give
$(\lambda^{a-z} A+Z)C=\bar C(\lambda^{a-\bar z} A+\bar Z)=0$. 
Recall that the VEV of $A$ is 
proportional to the $B-L$ generator $Q_{B-L}$ as 
$\VEV{A}=\frac{3}{2}vQ_{B-L}$.
Also $C$, ${\bf 16}$, is decomposed into 
$({\bf 3},{\bf 2},{\bf 1})_{1/3}$, 
$({\bf \bar 3},{\bf 1},{\bf 2})_{-1/3}$, 
$({\bf 1},{\bf 2},{\bf 1})_{-1}$ and $({\bf 1},{\bf 1},{\bf 2})_{1}$ 
under
$SU(3)_C\times SU(2)_L\times SU(2)_R\times U(1)_{B-L}$.
Since $\VEV{\bar CC}\neq 0$, 
not all components in the spinor $C$ vanish. 
Then $Z$ is fixed to be $Z\sim -\frac{3}{2}\lambda v Q_{B-L}^0$, 
where $Q_{B-L}^0$ is
the $B-L$ charge of the component field in $C$, which has non-vanishing VEV. 
It is interesting
that no other component fields can have non-vanishing VEVs 
because of the $F$-flatness
conditions. If  the $({\bf 1},{\bf 1},{\bf 2})_1$ field obtains
a non-zero VEV (therefore, $\VEV{Z}\sim -\frac{3}{2}\lambda v$), then the 
gauge group 
$SU(3)_C\times SU(2)_L\times SU(2)_R\times U(1)_{B-L}$ is broken to the 
standard gauge group. Once the direction of the VEV $\VEV{C}$ is 
determined, the VEV $\VEV{\bar C}$ must have the same direction 
because of the $D$-flatness
condition. Therefore, $\VEV{\bar Z}\sim -\frac{3}{2}\lambda v$.
Thus, all VEVs have now been fixed.

Finally the $F$-flatness condition of $H^\prime$ is examined. 
$W_{H^\prime}$, which is linear in the $H^\prime$ field, is
written
\begin{equation}
W_{H^\prime}=\lambda^{h+a+h^\prime}H^\prime AH. 
\end{equation}
The $F_{H^\prime}$ leads to the vanishing VEV of the triplet
Higgs $\VEV{H_T}=0$. 

There are several terms which must be forbidden for the stability of the 
DW mechanism. For example, $H^2$, $HZH^\prime$ and $H\bar Z H^\prime$
induce a large mass of the doublet Higgs, 
and the term $\bar CA^\prime A C$ would destabilize the DW form of 
$\VEV{A}$.
We can easily forbid these terms using the SUSY zero mechanism.
For example, if we choose
$h<0$, then $H^2$ is forbidden, and if we choose $\bar c+c+a+a^\prime<0$, 
then
$\bar CA^\prime A C$ is forbidden. 
Once these dangerous terms are forbidden
by the SUSY zero mechanism, higher-dimensional terms which also become
dangerous;
for example, 
$\bar CA^\prime A^3 C$ and $\bar CA^\prime C\bar CA C$ are automatically
forbidden, since only gauge invariant operators with negative charge
can have non-vanishing VEVs. This is also an attractive point of our 
scenario. 

In the end of this section, we would like to explain how to determine
the symmetry and the quantum numbers in the Higgs sector to realize 
DT splitting.
It is essential that the dangerous terms are forbidden by SUSY zero mechanism
and the necessary terms must be allowed by the symmetry. The dangerous terms
are
\begin{equation}
H^2, HH',HZH', \bar CA'C, \bar CA'AC, \bar CA'ZC, A'A^4, A'A^5.
\end{equation}
On the other hand, the terms required to realize DT splitting well are
\begin{equation}
A'A, A'A^3, HAH',\bar C'(A+Z)C, \bar C(A+Z)C', S\bar CC.
\end{equation}
Here we denote both $Z$ and $\bar Z$ as $Z$.
In order to forbid $HH'$ while $HAH'$ is allowed, we introduce $Z_2$ parity.
We have some ambiguities to assign the $Z_2$ parity, but once the parity is 
fixed, the above requirements become just inequalities, which are easily
satisfied as discussed in this section.

Of course, the above conditions are necessary but not sufficient. 
As in the next section, we have to write down the mass matrices of 
Higgs sector to know whether an assignment truly works well or not.

\section{Mass spectrum of Higgs sector}
In this section, we examine the mass spectrum of the super-heavy particles. 
Before going to the detail, we classify the fields by the quantum number
of the standard gauge group.
Using the definition of the fields $Q({\bf 3,2})_{\frac{1}{6}}$,
$U^c({\bf \bar 3,1})_{-\frac{2}{3}}$, $D^c({\bf \bar 3,1})_{\frac{1}{3}}$,
$L({\bf 1,2})_{-\frac{1}{2}}$, $E^c({\bf 1,1})_1,N^c({\bf 1,1})_0$,
$X({\bf 3,2})_{-\frac{5}{6}}$ and their conjugate fields, and
$G({\bf 8,1})_0$ and $W({\bf 1,3})_0$ with the standard gauge symmetry,
under $SO(10)\supset SU(5) \supset SU(3)_C\times SU(2)_L\times U(1)_Y$,
the spinor ${\bf 16}$, vector ${\bf 10}$ and the adjoint ${\bf 45}$
are
\begin{eqnarray}
{\bf 16}&\rightarrow &
\underbrace{[Q+U^c+E^c]}_{\bf 10}+\underbrace{[D^c+L]}_{\bf \bar 5}
+\underbrace{N^c}_{\bf 1},\\
{\bf 10}&\rightarrow &
\underbrace{[D^c+L]}_{\bf \bar 5}+\underbrace{[\bar D^c+\bar L]}_{\bf 5},\\
{\bf 45}&\rightarrow &
\underbrace{[G+W+X+\bar X+N^c]}_{\bf 24}
+\underbrace{[Q+U^c+E^c]}_{\bf 10}
+\underbrace{[\bar Q+\bar U^c+\bar E^c]}_{\bf \overline{10}}
+\underbrace{N^c}_{\bf 1}.
\end{eqnarray}

In the followings, we study how mass matrices of the above fields
are determined by anomalous $U(1)_A$ charges.
For the mass terms, we must
take account of not only the terms in the previous section but also
the terms that contain two fields with vanishing
VEVs. 

First we examine the mass spectrum of ${\bf 5}$ and ${\bf \bar 5}$ of
$SU(5)$. 
Considering the additional terms
$\lambda^{2h^\prime} H^\prime H^\prime$,
$\lambda^{c^\prime+\bar c^\prime}\bar C^\prime C^\prime$, 
$\lambda^{c'+c+h'}C'CH'$,
$\lambda^{\bar c'+\bar c+h'}\bar C'\bar CH'$ and
$\lambda^{2\bar c+h'}\bar C^2H'$,
we write the mass matrices $M_I$,
which are for the representations $I=D^c(H_T),L(H_D)$ and their 
conjugates:
\begin{equation}
M_I=\bordermatrix{
       &I_H & I_{H^\prime} & I_{C} &I_{C^\prime} \cr
\bar I_H & 0 & \lambda^{h+h^\prime +a}\VEV{A} & 0 & 0 \cr
\bar I_{H^\prime} & \lambda^{h+h^\prime +a}\VEV{A} & \lambda^{2h^\prime} 
& 0 & \lambda^{h'+c'+c}\VEV{C} \cr
\bar I_{\bar C}& 0 & \lambda^{h'+2\bar c}\VEV{\bar C} & 0 
& \lambda^{\bar c+c^\prime+a}\beta_Iv \cr
\bar I_{\bar C^\prime} &  0 & \lambda^{h'+\bar c'+\bar c}\VEV{\bar C}
 & \lambda^{c+\bar c^\prime+a}\beta_Iv 
  & \lambda^{c^\prime+\bar c^\prime} \cr}.
\end{equation}

The colored Higgs obtain their masses of order 
$\lambda^{h+h^\prime+a}\VEV{A}\sim \lambda^{h+h^\prime}$.
Since in general $\lambda^{h+h^\prime}>\lambda^{2h^\prime}$,
the proton decay is naturally suppressed. The effective colored
Higgs mass is estimated as
$(\lambda^{h+h^\prime})^2/\lambda^{2h^\prime}=\lambda^{2h}$, 
which is larger than the cutoff scale, because
$h<0$.
One pair of the doublet Higgs is massless,
while another pair of doublet Higgs acquires a mass of order 
$\lambda^{2h^\prime}$.
The DW mechanism works well, although we have to examine the effect of the 
rather
light super-heavy particles.
Since $\beta_{D^c}=-2$ and $\beta_L=-3$, the color 
triplets acquire masses $2\lambda^{\bar c+c^\prime}$ and 
$2\lambda^{c+\bar c^\prime}$, while the weak doublets acquire masses  
$3\lambda^{\bar c+c^\prime}$ and 
$3\lambda^{c+\bar c^\prime}$.

Note that if the term $\bar C'A\bar CH$, which is not allowed with the typical 
charge assignment in Table I, is allowed by the symmetry, the
massless Higgs doublet becomes
\begin{equation}
\bar 5_H+\lambda^{h-c+\frac{1}{2}(\bar c-c)}\bar 5_C,
\label{mixing}
\end{equation}
and the effect of the mixing must
be taken into account in considering the quark and lepton mass matrices.

Next we examine the mass matrices for the representations 
$I=Q,U^c$ and $E^c$,
which are contained in the {\bf 10} of $SU(5)$.
Like the superpotential previously discussed, the additional terms
$\lambda^{2a^\prime}A^\prime A^\prime$, 
$\lambda^{c^\prime+\bar c^\prime}\bar C^\prime C^\prime$,
$\lambda^{c^\prime+a^\prime+\bar c} \bar CA^\prime C^\prime$ and
$\lambda^{\bar c^\prime+a^\prime+c} \bar C^\prime A^\prime C$
must be taken into account.
The mass matrices are written as $4\times 4$ matrices,
\begin{equation}
M_I=\bordermatrix{
        &I_A & I_{A^\prime} &I_{ C}&I_{C^\prime} \cr
\bar I_A &0& \lambda^{a^\prime+a} \alpha_I & 0  & 
\frac{\lambda^{\bar c+c^\prime+a}}{\sqrt{2}}\VEV{\bar C} \cr
\bar I_{A^\prime} &\lambda^{a+a^\prime} \alpha_I & \lambda^{2a^\prime} & 0 & 
\frac{\lambda^{\bar c+c^\prime+a^\prime}}{\sqrt{2}}\VEV{\bar C} \cr
\bar I_{\bar C}&0 & 0 & 0 & \lambda^{\bar c+c^\prime+a}\beta_Iv \cr
\bar I_{\bar C^\prime} &\frac{\lambda^{c+\bar c^\prime+a}}{\sqrt{2}}\VEV{C} &
\frac{\lambda^{c+\bar c^\prime+a^\prime}}{\sqrt{2}}\VEV{C} &
\lambda^{c+\bar c^\prime+a}\beta_Iv & \lambda^{c^\prime+\bar c^\prime}\cr}.
\label{mass10}
\end{equation}
where $\alpha_I$ vanishes for $I=Q$ and $U^c$ because
these are Nambu-Goldstone modes, but $\alpha_{E^c}\neq 0$.
On the other hand, $\beta_I=\frac{3}{2}((B-L)_I-1)$; that is, $\beta_Q=-1$, 
$\beta_{U^c}=-2$ and $\beta_{E^c}=0$. 
Thus for each $I$, the $4\times 4$ matrix has one 
vanishing eigenvalue, which corresponds to the Nambu-Goldstone mode eaten
by the Higgs mechanism. The mass spectrum of the remaining three
modes is ($\lambda^{c+\bar c^\prime+a}v$, $\lambda^{c^\prime+\bar c+a}v$,
$\lambda^{2a^\prime}$) for the color-triplet modes $Q$ and $U^c$, and
($\lambda^{a+a^\prime}$, 
$\lambda^{a+a^\prime}$,
$\lambda^{c^\prime+\bar c^\prime}$) or 
($\lambda^{c+\bar c^\prime+a}\VEV{C}$, 
$\lambda^{c^\prime+\bar c+a}\VEV{\bar C}$,
$\lambda^{2a^\prime}$) for the color-singlet modes $E^c$.

The adjoint fields $A$ and $A^\prime$ contain
two $G$, two $W$ and two pairs of $X$ and $\bar X$, 
whose mass matrices $M_I(I=G,W,X)$ is given by
\begin{equation}
M_I=\bordermatrix{
            &    I_A       &        I_{A'}           \cr
\bar I_A    &     0        & \alpha_I\lambda^{a+a'}  \cr
\bar I_{A'} & \alpha_I\lambda^{a+a'} & \lambda^{2a'}  \cr}.
\end{equation}
Two $G$ and two $W$ acquire masses $\lambda^{a^\prime+a}$.
Since $\alpha_X=0$, one pair of $X$ is massless, which is eaten
by Higgs mechanism. However, the other pair has a rather light mass of
$\lambda^{2a^\prime}$.

Once we determine the anomalous $U(1)_A$ charges, the mass spectrum
of all fields is determined, and hence we can examine whether 
the running couplings from the low energy scale meet at the unification
scale or not. Before going to the discussion of the conditions
for gauge coupling unification, in the next section,
we will examine several models with this DT splitting mechanism
and conditions with which realistic mass matrices of quarks and leptons
can be obtained.

\section{Quark and Lepton masses and Neutrino relation}
In this section, we examine models to demonstrate 
how to determine
everything from the anomalous $U(1)_A$ charges.

In addition to the Higgs sector in Table.I, we introduce only three 
${\bf 16}$ representations $\Psi_i$
with anomalous $U(1)_A$ charges $(\psi_1=n+3,\psi_2=n+2,\psi_3=n)$ 
and one ${\bf 10}$ field
$T$ with
charge $t$ as the matter contents. These matter fields are assigned
odd R-parity, while those of the Higgs sector are assigned even 
R-parity.
Such an assignment of R-parity guarantees that the argument regarding
VEVs 
in the previous section does not change if these matter fields have 
vanishing VEVs.
Then the mass term of ${\bf 5}$ and ${\bf \bar 5}$ of $SU(5)$
is written 
\begin{eqnarray}
&&{\bf 5}_T ( \lambda^{t+\psi_1+c} \VEV{C}, \lambda^{t+\psi_2+c}\VEV{C}, 
\lambda^{t+\psi_3+c}\VEV{C}, \lambda^{2t})
\left(
\begin{array}{c}  {\bf \bar 5}_{\Psi1} \\ {\bf \bar 5}_{\Psi2} \\ 
{\bf \bar 5}_{\Psi3}
 \\ {\bf \bar 5_T}
\end{array}
\right) \\
&=&{\bf 5}_T ( \lambda^{t+\psi_1+(c-\bar c)/2}, 
               \lambda^{t+\psi_2+(c-\bar c)/2}, 
               \lambda^{t+\psi_3+(c-\bar c)/2}, \lambda^{2t})
\left(
\begin{array}{c}  {\bf \bar 5}_{\Psi1} \\ {\bf \bar 5}_{\Psi2} \\ 
{\bf \bar 5}_{\Psi3}
 \\ {\bf \bar 5_T}
\end{array}
\right),
\end{eqnarray}
where $\VEV{\bar C}=\VEV{C}\sim \lambda^{-(c+\bar c)/2}$.
Since $\psi_3<\psi_2<\psi_1$, 
the massive mode ${\bf \bar 5}_M$, the partner of $5_T$, must
be ${\bf \bar 5}_{\Psi3}(\Delta\equiv 2t-(t+\psi_3+(c-\bar c)/2)>0)$ or
${\bf \bar 5}_T(\Delta<0)$. 
The former case is interesting and the massive mode
is given by
\begin{equation}
{\bf \bar 5}_M \sim {\bf \bar 5}_{\Psi3}+\lambda^{\Delta}{\bf \bar 5}_T
                    +\lambda^{\psi_2-\psi_3}{\bf \bar 5}_{\Psi2}
                    +\lambda^{\psi_1-\psi_3}{\bf \bar 5}_{\Psi1}.
\end{equation} 
Therefore the three massless modes 
$({\bf \bar 5}_1, {\bf \bar 5}_2, {\bf \bar 5}_3) $ 
are
written $({\bf \bar 5}_{\Psi1}+\lambda^{\psi_1-\psi_3}{\bf \bar 5}_{\Psi3}, 
{\bf \bar 5}_T+ \lambda^{\Delta}{\bf \bar 5}_{\Psi3},
{\bf \bar 5}_{\Psi2}+\lambda^{\psi_2-\psi_3}{\bf \bar 5}_{\Psi3})$. 
The Dirac mass matrices for quarks and leptons 
can be 
obtained from the interaction
\begin{equation}
\lambda^{\psi_i+\psi_j+h}\Psi_i\Psi_jH.
\label{Yukawa}
\end{equation}
The mass matrices for the up quark sector and the down quark sector are
\begin{equation}
M_u=\left(
\begin{array}{ccc}
\lambda^6 & \lambda^5 & \lambda^3 \\
\lambda^5 & \lambda^4 & \lambda^2 \\
\lambda^3 & \lambda^2 & 1    
\end{array}
\right)\VEV{H_u},\quad
M_d=\lambda^2\left(
\begin{array}{ccc}
\lambda^4 & \lambda^{\Delta+1} & \lambda^3 \\
\lambda^3 & \lambda^{\Delta} & \lambda^2 \\
\lambda^1 & \lambda^{\Delta-2} & 1    
\end{array}
\right)\VEV{H_d}.
\label{quark}
\end{equation}
Here taking $1\leq \Delta \leq 3$ leads reasonable value of 
the ratio $m_s/m_b$. 
Note that the Yukawa couplings for 
${\bf \bar 5}_2\sim {\bf \bar 5}_T+\lambda^{\Delta}{\bf \bar 5}_{\Psi 3}$
are obtained only through the Yukawa couplings for the component 
${\bf \bar 5}_{\Psi 3}$,
because we have no Yukawa couplings for $T$ without the Higgs mixing 
(\ref{mixing}).
With the  Higgs mixing (\ref{mixing}), the interaction
$\lambda^{\psi_i+t+c}\Psi_iTC$ induces the correction to the 
mass matrix of down-type quarks. 
It is easily checked that the correction to the down-type Yukawa couplings
are the same order as in Eq. (\ref{quark}).

We can estimate the Cabbibo-Kobayashi-Maskawa (CKM) matrix\cite{CKM}
 from these quark matrices\footnote{
Strictly speaking, if the Yukawa coupling originated only from the 
interaction (\ref{Yukawa}), the mixing concerning to the first generation
becomes too small 
because of a cancellation. In order to get the expected value of
CKM matrix as in Eq. (\ref{CKM}), non-renormalizable terms, for
example, $\Psi_i\Psi_jH\bar CC$ must be taken into account.
It is required that $c+\bar c\geq -5$.
}
 as
\begin{equation}
U_{\rm CKM}=
\left(
\begin{array}{ccc}
1 & \lambda &  \lambda^3 \\
\lambda & 1 & \lambda^2 \\
\lambda^3 & \lambda^2 & 1
\end{array}
\right),
\label{CKM}
\end{equation}
which is consistent with the experimental value if we choose 
$\lambda\sim 0.2$.
Since the ratio of the Yukawa couplings of top and bottom quarks is 
$\lambda^2$,
a small value of $\tan \beta\equiv \VEV{H_u}/\VEV{H_d}$ is 
predicted by these mass matrices.

The Yukawa matrix for the charged lepton sector is the same as the transpose 
of $M_d$ at this stage, except for an overall factor $\eta$ induced by the 
renormalization group effect. 
The mass matrix for the Dirac mass of neutrinos is given by
\begin{equation}
M_{\nu_D}=\lambda^2\left(
\begin{array}{ccc}
\lambda^4 & \lambda^3 & \lambda \\
\lambda^{\Delta+1} & \lambda^{\Delta} & \lambda^{\Delta-2} \\
\lambda^3   & \lambda^2         & 1 
\end{array}
\right)\VEV{H_u}\eta.
\end{equation}

The right-handed neutrino masses come from the interaction
\begin{equation}
\lambda^{\psi_i+\psi_j+2\bar c}\Psi_i\Psi_j\bar C\bar C
\end{equation}
as
\begin{equation}
M_R=\lambda^{\psi_i+\psi_j+2\bar c}\VEV{\bar C}^2
=\lambda^{2n+\bar c-c}\left(
\begin{array}{ccc}
\lambda^6 & \lambda^5 & \lambda^3 \\
\lambda^5 & \lambda^4 & \lambda^2 \\
\lambda^3   & \lambda^2         & 1
\end{array}
\right).
\end{equation}
Therefore we can estimate the neutrino mass matrix:
\begin{equation}
M_\nu=M_{\nu_D}M_R^{-1}M_{\nu_D}^T=\lambda^{4-2n+c-\bar c}\left(
\begin{array}{ccc}
\lambda^2          & \lambda^{\Delta-1}  & \lambda \\
\lambda^{\Delta-1} & \lambda^{2\Delta-4} & \lambda^{\Delta-2} \\
\lambda            & \lambda^{\Delta-2}  & 1 
\end{array}
\right)\VEV{H_u}^2\eta^2.
\end{equation}
Note that the overall factor $\lambda^{4-2n+c-\bar c}$ can
have negative power.
From these mass matrices in the lepton sector the
Maki-Nakagawa-Sakata (MNS)
matrix is obtained as
\begin{equation}
U_{\rm MNS}=
\left(
\begin{array}{ccc}
1 & \lambda^{3-\Delta} &  \lambda \\
\lambda^{3-\Delta} & 1 & \lambda^{\Delta-2} \\
\lambda & \lambda^{\Delta-2} & 1
\end{array}
\right)
\end{equation}
for $2\leq \Delta \leq 3$ and
\begin{equation}
U_{\rm MNS}=
\left(
\begin{array}{ccc}
1 & \lambda &  \lambda^{3-\Delta} \\
\lambda & 1 & \lambda^{2-\Delta} \\
\lambda^{3-\Delta} & \lambda^{2-\Delta} & 1
\end{array}
\right)
\end{equation}
for $1\leq \Delta \leq 2$.
If we take $\Delta=5/2$, namely,
\begin{equation}
t=n+\frac{1}{2}(c-\bar c+5),
\label{neutrino1}
\end{equation}
bi-large neutrino mixing angle
is obtained. 
We then obtain the prediction 
$m_{\nu_\mu}/m_{\nu_\tau}\sim \lambda$, which is consistent with
the experimental data: 
$1.6\times 10^{-3} {\rm eV}^2\leq \Delta m_{\rm atm}^2\leq 4
\times 10^{-3}
{\rm eV}^2$
and $2\times 10^{-5} {\rm eV}^2\leq \Delta m_{\rm solar}^2\leq 1
\times 10^{-4}
{\rm eV}^2$ (LMA).
The relation $V_{e3}\sim \lambda$ is also an interesting 
prediction from this 
matrix, though CHOOZ gives a restrictive upper limit $V_{e3}\leq 0.15$.
\cite{CHOOZ}
Moreover, if we take
$4-2n+c-\bar c=-(5+l)$, the parameter
$l$ is determined from
\begin{equation}
\lambda^l=\lambda^{-5}\frac{H_u^2\eta^2}{m_{\nu_\tau}\Lambda},
\end{equation}
where $m_{\nu_\tau}$ is tau neutrino mass.
We are supposing that the cutoff scale $\Lambda$ is in a range
$10^{16}({\rm GeV})<\Lambda<10^{20}({\rm GeV})$, which allows
us to take $-2\leq l \leq 2$. 
If we take $l=0$,
the neutrino masses are given  by
$m_{\nu_\tau}\sim \lambda^{-5}\VEV{H({\bf 10,5})}^2\eta^2/\Lambda\sim
m_{\nu_\mu}/\lambda
\sim m_{\nu_e}/\lambda^2$. If we take $\eta\VEV{H({\bf 10,5})}=100$ GeV,
$\Lambda\sim 10^{18}$ GeV and $\lambda=0.2$, then we get
$m_{\nu_\tau}\sim 3\times 10^{-2}$ eV, $m_{\nu_\mu}\sim 6\times
10^{-3}$ eV
and $m_{\nu_e}\sim 1\times 10^{-3}$ eV. From such a rough estimation, 
we can obtain almost
desirable values for explaining
the experimental data from the atmospheric neutrino and large
mixing angle (LMA) MSW solution for solar neutrino 
problem.\cite{MSW} This LMA solution for the solar neutrino problem
gives the best fitting to the present experimental data.\cite{Valle}
\footnote{
If we take $\Delta=2$, namely $t=n+\frac{1}{2}(c-\bar c+4)$, 
the MNS matrix becomes lopsided type. It has been argued that
even in this case, the desirable values can be obtained, using 
the ambiguity of coefficients.\cite{lopsided}
}

In addition to Eq.~(\ref{Yukawa}), the interactions
\begin{equation}
\lambda^{\psi_i+\psi_j+2a+h}\Psi_iA^2\Psi_jH
\end{equation}
also contribute to the Yukawa couplings. Here $A$ is squared because
it has odd parity.
Since $A$ is proportional to the generator of $B-L$,
the contribution to the lepton Yukawa coupling is nine times larger
than that to quark Yukawa coupling, which can change the unrealistic
prediction $m_\mu=m_s$ at the GUT scale. 
Since the prediction $m_s/m_b\sim \lambda^{5/2}$ at the GUT scale is 
consistent with experiment, 
the enhancement factor $2\sim 3$ of $m_\mu$ can improve the situation.
Note that the additional terms contribute mainly in the lepton
sector.
If we set $a=-2$,
the additional matrices are
\begin{eqnarray}
\frac{\Delta M_u}{\VEV{H_u}}&=&
\frac{v^2}{4}\left(
\begin{array}{ccc}
\lambda^2 & \lambda & 0 \\
\lambda & 1 & 0 \\
0 & 0 & 0
\end{array}
\right)
,\quad
\frac{\Delta M_d}{\VEV{H_d}}=
\frac{v^2}{4}\left(
\begin{array}{ccc}
\lambda^2 & 0 & \lambda \\
\lambda & 0 & 1 \\
0 & 0 & 0 
\end{array}
\right)
,
\\
\frac{\Delta M_e}{\VEV{H_d}}&=&
\frac{9v^2}{4}\left(
\begin{array}{ccc}
\lambda^2 & \lambda & 0 \\
0 & 0 & 0 \\
\lambda & 1 & 0 
\end{array}
\right).
\end{eqnarray}
It is interesting that this modification essentially changes the 
eigenvalues of
only the first and second generation. Therefore it is natural to expect 
that a realistic mass pattern can be obtained by this modification.
This is one of the largest motivations to choose $a=-2$.
Note that this charge assignment also determines the scale 
$\VEV{A}\sim \lambda^2$.
It is suggestive that the fact that the GUT scale is 
slightly
smaller than the Planck scale is correlated with the discrepancy between
the naive prediction of the ratio $m_\mu/m_s$ from the unification and 
the experimental value. 
It is also interesting that the SUSY zero mechanism plays an essential 
role again. 
When $z, \bar z \geq -4$, the terms 
$\lambda^{\psi_i+\psi_j+a+z+h}Z\Psi_iA\Psi_jH+
\lambda^{\psi_i+\psi_j+2z+h}Z^2\Psi_i\Psi_jH
$ also contribute to the fermion mass matrices, though only to the first
generation.
It is useful to examine other charge assignment to $a$. 
If $a\leq -3$, then the modification changes the eigenvalue of at the most 
first 
generation, which is inconsistent with the present experimental results.
If $a=-1$, then the modification changes the eigenvalues of all generations.
It is consistent with the present experimental values,
though it does not explain the Yukawa coupling relation $y_b=y_\tau$
at the GUT scale. Since the GUT relation $y_b=y_\tau$ is still 
controvertible\cite{yamaguchi}, this option $a=-1$ may be realistic.

Proton decay mediated by the colored Higgs is strongly suppressed
in this model. As mentioned in the previous section, the effective 
mass of
the colored Higgs is of order $\lambda^{2h}\sim \lambda^{-12}$, which is 
much larger than the cutoff scale. 
Proton decay is also induced by the non-renormalizable
term
\begin{equation} 
\lambda^{\psi_i+\psi_j+\psi_k+\psi_l}\Psi_i\Psi_j\Psi_k\Psi_l,
\end{equation}
which has also the same suppression as via the colored Higgs mediation.

\section{A natural solution for the $\mu$ problem}
In our scenario, SUSY zero mechanism forbids the SUSY Higgs mass term
$\mu HH$. However, once SUSY is broken, the Higgs mass $\mu$ must be
induced. The induced mass must be proportional to the SUSY breaking scale.

We now examine a solution for the $\mu$ problem in a simple example
\cite{maekawa2}.
The essential point of this mechanism is that the VEV shift of a heavy
singlet field by SUSY breaking. 
We introduce the superpotential $W=\lambda^{s'}S'+\lambda^{s'+p}S'P$, where 
$S'$ and $P$ are singlet fields with positive anomalous $U(1)_A$ charge $s$ 
and with negative charge $p$, respectively ($s'+p\geq 0$). 
Note that the single term of $P$
is not allowed by SUSY zero mechanism, while usual symmetry cannot forbid
this term. This is an essential point of this mechanism. The SUSY vacuum is 
at
$\VEV{S'}=0$ and $\VEV{P}=\lambda^{-p}$. After SUSY is broken, these 
VEVs are modified. To determine the VEV shift of $S'$, which we would 
like to know because the singlet $S'$ with positive charge can couple to 
the Higgs field with negative charge, the most important SUSY breaking 
term is the tadpole term of $S'$, namely $\lambda^{s'} \Mp^2A S'$. Here $A$ 
is a SUSY breaking parameter of order of the weak scale. By this tadpole 
term, the VEV of $S'$ appears as $\VEV{S'}=\lambda^{-s'-2p}A$. If we have 
$\lambda^{s'+2h} S'H^2$, the SUSY Higgs mass is obtained as 
$\mu=\lambda^{2h-2p}m_{SB}$, which is proportional to the SUSY breaking 
parameter $m_{SB}$ and the proportional coefficient can be
of order 1 if $h\sim p$. Note that the $F$-term of
$S'$ is calculated as $F_{S'}\sim \lambda^{-s'-2p}m_{SB}^2$.
 The Higgs mixing term $B\mu$ can be obtained from the SUSY term
$\lambda^{s'+2h}S'H^2$ and the
SUSY breaking term $\lambda^{s'+2h}A_{S'H^2}S'H^2$ as
$\lambda^{s'+2h}F_{S'}\sim \lambda^{2h-2p}m_{SB}^2$ and 
$\lambda^{2h-2p}A^2\sim \mu A$, respectively.
Therefore the relation $B\sim m_{SB}$ is naturally obtained
\footnote{
If doublet-triplet splitting is realized by fine-tuning or some
accidental cancellation, the Higgs mixing
$B\mu$ can become intermediated scale $m_{SB} M_X$ as discussed in
Ref.~\cite{kawamura}, where $M_X$ is the GUT scale.
However, once the 
doublet-triplet splitting is naturally solved  as in 
Ref.~\cite{maekawa}, such a problem disappears }.
This is a solution for the $\mu$ problem. Note that the condition 
$h\sim p$ can be satisfied because both fields $H$ and $P$ have
 negative charges. 
Note that $S'$ or $P$ can be a composite operator, for example,
a composite operator $\bar CC$ can play the same role as $P$ in
the above mechanism. In this case, the condition becomes
\begin{equation}
h\sim p=\frac{1}{2}(c+\bar c).
\end{equation}
We call this condition the economical condition for the $\mu$ problem.

\section{Conditions for gauge coupling unification}
In order to stabilize the DW form of $\VEV{A}$, the term
$\bar C A^\prime A C$ must be forbidden by SUSY zero mechanism,
namely, $\bar c +c+a^\prime + a<0$. On the other hand, $a^\prime+3a\geq 0$
is required to obtain the term $A^\prime A^3$. From these inequalities,
we obtain $\frac{1}{2}(c+\bar c)<a$, which leads to
$\VEV{A}\sim \lambda^{-a}>\lambda^{-(c+\bar c)/2}\sim \VEV{C}=\VEV{\bar C}$. 
Therefore at the scale $\Lambda_A\equiv\VEV{A}\sim \lambda^{-a}$, 
$SO(10)$ gauge group is broken into
$SU(3)_C\times SU(2)_L\times SU(2)_R\times U(1)_{B-L}$, which is broken
into the standard gauge group $SU(3)_C\times SU(2)_L\times U(1)_Y$ 
at the scale $\Lambda_C\equiv\VEV{C}\sim \lambda^{-(c+\bar c)/2}$. 

In this paper, we make an analysis based on the renormalization group 
equations up to one loop. 
The conditions of the gauge coupling unification are given by
\begin{equation}
\alpha_3(\Lambda_A)=\alpha_2(\Lambda_A)=
\frac{3}{5}\alpha_Y(\Lambda_A)\equiv\alpha_1(\Lambda_A),
\end{equation}
where 
$\alpha_1^{-1}(\mu>\Lambda_C)\equiv 
\frac{3}{5}\alpha_R^{-1}(\mu>\Lambda_C)
+\frac{2}{5}\alpha_{B-L}^{-1}(\mu>\Lambda_C)$.
Here $\alpha_X=\frac{g_X^2}{4\pi}$ and 
the parameters $g_X (X=3,2,R,B-L,Y)$ are the gauge couplings of 
$SU(3)_C$, $SU(2)_L$, $SU(2)_R$, $U(1)_{B-L}$ and $U(1)_Y$, 
respectively.

The gauge couplings at the scale $\Lambda_A$ are roughly described by
\begin{eqnarray}
\alpha_1^{-1}(\Lambda_A)&=&\alpha_1^{-1}(M_{SB})
+\frac{1}{2\pi}\left(b_1\ln \left(\frac{M_{SB}}{\Lambda_A}\right)
+\Sigma_i \Delta b_{1i}\ln \left(\frac{m_i}{\Lambda_A}\right)
-\frac{12}{5}\ln \left(\frac{\Lambda_C}{\Lambda_A}\right)\right), 
\label{alpha1}\\
\alpha_2^{-1}(\Lambda_A)&=&\alpha_2^{-1}(M_{SB})
+\frac{1}{2\pi}\left(b_2\ln \left(\frac{M_{SB}}{\Lambda_A}\right)
+\Sigma_i \Delta b_{2i}\ln \left(\frac{m_i}{\Lambda_A}\right)\right), \\
\alpha_3^{-1}(\Lambda_A)&=&\alpha_3^{-1}(M_{SB})
+\frac{1}{2\pi}\left(b_3\ln \left(\frac{M_{SB}}{\Lambda_A}\right)
+\Sigma_i \Delta b_{3i}\ln \left(\frac{m_i}{\Lambda_U}\right)\right), 
\end{eqnarray}
where $M_{SB}$ is a SUSY breaking scale, 
$(b_1,b_2,b_3)=(33/5,1,-3)$ are the 
renormalization group coefficients
for the minimal SUSY standard model(MSSM)
and $\Delta b_{ai}(a=1,2,3)$ are the correction to the coefficients 
from the massive fields with mass $m_i$.
The last term in Eq. (\ref{alpha1}) is from the breaking 
$SU(2)_R\times U(1)_{B-L}\rightarrow U(1)_Y$ by the VEV $\VEV{C}$.
Since the gauge couplings at the SUSY breaking scale $M_{SB}$
are given by
\begin{eqnarray}
\alpha_1^{-1}(M_{SB})&=&\alpha_G^{-1}(\Lambda_G)
+\frac{1}{2\pi}\left(b_1\ln \left(\frac{\Lambda_G}{M_{SB}}\right)\right),\\
\alpha_2^{-1}(M_{SB})&=&\alpha_G^{-1}(\Lambda_G)
+\frac{1}{2\pi}\left(b_2\ln \left(\frac{\Lambda_G}{M_{SB}}\right)\right), \\
\alpha_3^{-1}(M_{SB})&=&\alpha_G^{-1}(\Lambda_G)
+\frac{1}{2\pi}\left(b_3\ln \left(\frac{\Lambda_G}{M_{SB}}\right)\right),
\end{eqnarray}
where $\alpha_G^{-1}(\Lambda_G)\sim 25$ and 
$\Lambda_G\sim 2\times 10^{16}$ GeV, the above conditions for unification
are rewritten as
\begin{eqnarray}
&&b_1\ln \left(\frac{\Lambda_A}{\Lambda_G}\right)
+\Sigma_I\Delta b_{1I}\ln \left(\frac{\Lambda_A^{\bar r_I}}{\det \bar M_I}
\right)
-\frac{12}{5}\ln \left(\frac{\Lambda_A}{\Lambda_C}\right) \\
&=&b_2\ln \left(\frac{\Lambda_A}{\Lambda_G}\right)
+\Sigma_I\Delta b_{2I}\ln\left(\frac{\Lambda_A^{\bar r_I}}{\det \bar M_I}
\right) \\
&=&b_3\ln \left(\frac{\Lambda_A}{\Lambda_G}\right)
+\Sigma_I\Delta b_{3I}\ln\left(\frac{\Lambda_A^{\bar r_I}}{\det \bar M_I}
\right),
\end{eqnarray}
where $\bar M_I$ are the reduced mass matrices which have no massless
mode and $\bar r_I$ are rank of the reduced mass matrices. For example,
\begin{equation}
\bar M_L=\bordermatrix{
        & L_{H^\prime} & L_{\bar C} &L_{\bar C^\prime} \cr
\bar L_{H^\prime} & \lambda^{2h^\prime} & 0 & 0 \cr
\bar L_{\bar C}& \lambda^{h'+2\bar c}\VEV{\bar C} & 0 & 
\lambda^{\bar c+c^\prime+a}\beta_Lv \cr
\bar L_{\bar C^\prime} & \lambda^{h'+\bar c'+\bar c}\VEV{\bar C}
 & \lambda^{c+\bar c^\prime+a}\beta_Lv 
  & \lambda^{c^\prime+\bar c^\prime} \cr}.
\end{equation}
The correction to the renormalization coefficients $\Delta b_{aI}$ are
given by
\begin{center}
\begin{tabular}{|c|c|c|c|c|c|c|c|c|}
\hline
$I$      & $Q+\bar Q$ & $U^c+\bar U^c$ & $E^c+\bar E^c$ & $D^c+\bar D^c$ 
         & $L+\bar L$ & $G $& $W$ & $X+\bar X$ \\
\hline
$\Delta b_{1I}$ & $\frac{1}{5}$& $\frac{8}{5}$& $\frac{6}{5}$& $\frac{2}{5}$
         & $\frac{3}{5}$ & 0 &  0 & 5 \\
\hline
$\Delta b_{2I}$ & 3 & 0 & 0 & 0 & 1 & 0 & 2 & 3 \\
\hline
$\Delta b_{3I}$ & 2 & 1 & 0 & 1 & 0 & 3 & 0 & 2 \\ 
\hline
\end{tabular}
\end{center}
In our scenario, the unification scale $\Lambda_A\sim \lambda^{-a}$, 
the symmetry breaking scale 
$\Lambda_C\sim \lambda^{-\frac{1}{2}(c+\bar c)}$ and the  determinants of 
the reduced mass matrices are fixed by the anomalous $U(1)_A$ charges;
\begin{eqnarray}
\det \bar M_{Q}&\sim &\det \bar M_{U^c}\sim 
\lambda^{2a'+c+\bar c+c'+\bar c'}, \label{detQ} \\
\det \bar M_{E^c}&\sim &\lambda^{2a+2a'+c'+\bar c'}, \\
\det M_{D^c}&\sim &\lambda^{2h+2h'+c+\bar c+c'+\bar c'}, \\
\det \bar M_{L}&\sim &\lambda^{2h'+c+\bar c+c'+\bar c'}, \\
\det M_{G}&\sim &\det M_W\sim \lambda^{2a+2a'}, \\
\det \bar M_{X}&\sim &\lambda^{2a'}. \label{detX}
\end{eqnarray}
The unification conditions $\alpha_1(\Lambda_A)=\alpha_2(\Lambda_A)$,
$\alpha_1(\Lambda_A)=\alpha_3(\Lambda_A)$ and
$\alpha_2(\Lambda_A)=\alpha_3(\Lambda_A)$ are rewritten as
\begin{eqnarray}
&&\left(\frac{\Lambda_A}{\Lambda_G}\right)^{14}
\left(\frac{\Lambda_C}{\Lambda_A}\right)^6
\left(\frac{\det \bar M_L}{\det \bar M_{D^c}}\right)
\left(\frac{\det \bar M_Q}{\det \bar M_{U}}\right)^4
\left(\frac{\det \bar M_Q}{\det \bar M_{E^c}}\right)^3
\left(\frac{\det \bar M_W}{\det \bar M_{X}}\right)^5  \\ \nonumber
&=& \Lambda_A^{-\bar r_{D^c}+\bar r_L-4\bar r_{U^c}-3\bar r_{E^c}+7\bar r_Q
-5\bar r_X+5\bar r_W}, \\
&&\left(\frac{\Lambda_A}{\Lambda_G}\right)^{16}
\left(\frac{\Lambda_C}{\Lambda_A}\right)^4
\left(\frac{\det \bar M_{D^c}}{\det \bar M_{L}}\right)
\left(\frac{\det \bar M_Q}{\det \bar M_{U}}\right)
\left(\frac{\det \bar M_Q}{\det \bar M_{E^c}}\right)^2
\left(\frac{\det \bar M_G}{\det \bar M_{X}}\right)^5  \\ \nonumber
&=&\Lambda_A^{-\bar r_{L}+\bar r_{D^c}-\bar r_{U^c}-2\bar r_{E^c}+3\bar r_Q
-5\bar r_X+5\bar r_G}, \\
&&\left(\frac{\Lambda_A}{\Lambda_G}\right)^{4}
\left(\frac{\det \bar M_{D^c}}{\det \bar M_{L}}\right)
\left(\frac{\det \bar M_U}{\det \bar M_{Q}}\right)
\left(\frac{\det \bar M_G}{\det \bar M_{W}}\right)^2
\left(\frac{\det \bar M_G}{\det \bar M_{X}}\right) \\ \nonumber
&=&\Lambda_A^{-\bar r_{L}+\bar r_{D^c}-\bar r_{Q}+\bar r_{U}-2\bar r_W
-\bar r_X+3\bar r_G}, 
\end{eqnarray}
which lead to
$\Lambda\sim \lambda^{\frac{h}{7}}\Lambda_G$,
$\Lambda\sim \lambda^{-\frac{h}{8}}\Lambda_G$ and 
$\Lambda\sim \lambda^{-\frac{h}{2}}\Lambda_G$, respectively.
So the unification condition becomes $h\sim 0$, and then 
the cutoff scale must be taken as $\Lambda\sim \Lambda_G$.
Note that these relation are independent on the anomalous $U(1)_A$ charges
except that of the doublet Higgs. It implies that this result can be applied
to rather general cases. 
On the other hand, we should not take this relation $h\sim 0$ seriously 
because we have an ambiguity 
of order one coefficients and use only one loop renormalization group 
equations. However, in order to catch the tendency, the above analysis
is fairly useful. 

Before going to the discussion of model buildings, it is useful to examine 
the reason to obtain the above result.
The essential point appears in estimating the ratio of determinants of mass
matrices between the components in the same multiplet of $SU(5)$ gauge 
group. 
Note that in Eqs. (\ref{detQ})$\sim $(\ref{detX}), the powers are given by
simple sums of the anomalous $U(1)_A$ charges. Therefore,
the ratio 
$\det \bar M_L/\det M_{D^c}$ can be easily estimated from the trivial relation
$\lambda^{2h}\det \bar M_L/\det M_{D^c}=1$, where $2h$ is the total charge
of massless modes (a pair of doublet Higgs fields). The ratio
$\det \bar M_Q/\det M_{E^c}$ is also determined by the relation
$\lambda^{2a}\det \bar M_Q/(\lambda^{c+\bar c}\det M_{E^c})=1$,
where $2a$ and $c+\bar c$ are the total charges of massless modes
(Nambu-Goldstone (NG) modes which appear by breaking 
$SO(10)\rightarrow SU(3)_C\times SU(2)_L\times SU(2)_R\times U(1)_{B-L}$
and $SU(2)_R\times U(1)_{B-L}\rightarrow U(1)_Y$, respectively. )
The ratio $\det \bar M_G/\det M_X$ is calculated by the relation
$\lambda^{2a}\det \bar M_G/\det M_X=1$, where $2a$ is the total charge
of massless modes (NG modes which appear by breaking
$SO(10)\rightarrow SU(3)_C\times SU(2)_L\times SU(2)_R\times U(1)_{B-L}$.).
 It is interesting that
all the effect of massless modes except Higgs doublet are cancelled out
in deriving the conditions for the gauge coupling unification.
It means that it is not accidental to realize the coupling unification
in our scenario though the cutoff scale becomes around 
$\Lambda_G\sim 2\times 10^{16}$
GeV and the unification scale becomes $\lambda^{-a}\Lambda_G$.
(So we cannot take $a\leq -2$, because of proton stability.)
Actually we have no solution to realize the coupling unification
if DT splitting does not happen (i.e., 
$(b_1,b_2,b_3)=(6,0,-3)$ and $\det M_L/\det M_{D^c}=1$).
Therefore this result is non-trivial and it is stimulating that 
the proton decay via dimension six operator may be seen in future 
because the unification scale must be smaller than the usual unification 
scale $\Lambda_G\sim 2\times 10^{16}$ GeV. 

From these estimation, it is obvious that 
the charges of massless modes are essential to examine whether
gauge couplings are unified at a scale or not. It is independent on
the detail of the contents of Higgs sector. Therefore we can 
formally examine the possibility of gauge coupling unification 
without building models of Higgs sector explicitly. 
For example, to examine the case in which other 
massless modes than one pair of Higgs doublet appear,
it is sufficient to take account of the anomalous $U(1)_A$ charges
of the massless modes.
 Unfortunately we could not find natural example
in which coupling unification of gauge couplings and DT splitting are
realized. 
For example,
if  we introduce an
additional adjoint field  with negative
charge, the additional massless modes $G$ and $W$ appear. The masses
are controlled by the SUSY breaking terms, so are expected to be around the
SUSY breaking scale. (Strictly speaking, we can compute the mass scale,
using the same mechanism for the $\mu$ problem as discussed in section 6.)
Then we can calculate the running flow of the gauge couplings and examine
whether coupling unification happens or not.
Since the ratios of the mass determinants are given by
\begin{eqnarray}
\frac{\lambda^{-a}\det \bar M_L}{\det M_{D^c}}&\sim &\lambda^{-2h-a}, \\
\frac{\det \bar M_{Q}}{\det \bar M_E}&\sim & \lambda^{c+\bar c-2a}, \\
\frac{m_G \det \bar M_{G}}{\lambda^{-a}\det \bar M_{X}}
&\sim &\lambda^{3a-2\Delta+\omega}, 
\end{eqnarray}
where $m_G=\lambda^\omega \Lambda$ is the mass of the massless mode of $G$
and $\Delta$ is the charge of the massless fields $G$ and $W$,
the above conditions for coupling unification become
\begin{eqnarray}
\alpha_2&=&\alpha_3\rightarrow \Lambda\sim 
\lambda^{-\frac{1}{4}(2h-2\Delta+\omega)}\Lambda_G, \\
\alpha_1&=&\alpha_2\rightarrow \Lambda\sim
\lambda^{-\frac{1}{14}(-2h-10\Delta+5\omega)}\Lambda_G.
\end{eqnarray}
These equations lead to unrealistic relation $2\Delta-\omega=-6h$.

In the next section, we will find out several models in which
the condition for the gauge coupling unification $h\sim 0$ is almost
satisfied.

\section{Some models}
In this section, we examine the cases in which all the fields become massive
except one pair of Higgs doublets. Then unification scale becomes
$\lambda^{-a}\Lambda_G$ as discussed in the previous section. 
So we should take $a=-1$ to stabilize
nucleon. Then $a'=3$ or 4 because the term $A'A^5$ must be forbidden and
the term $A'A^3$ is required. 
The unification condition is $h\sim 0$, but we have to take
negative $h$ to forbid the Higgs mass term $H^2$. Therefore
we would like to know how large negative charge $h=-2n$ can be adopted
in our scenario. 

Necessary conditions for realizing DT splitting and bi-large neutrino mixing
$(\Delta=5/2)$ are
\begin{eqnarray}
&&c-\bar c=2n-9-l, \\
&&t=n+\frac{1}{2}(c-\bar c+5), \label{t} \\
&&n+t+c\geq 0, \\
&&t\geq 0, \\
&&c+\bar c + a'+a<0.
\end{eqnarray}
The third and forth conditions are required because the terms
$\Psi_3TC$ and $T^2$ are needed in our scenario.
Since the cutoff scale $\Lambda\sim \Lambda_G=2\times 10^{16}$ GeV, 
we must adopt $l= -1$ or $-2$ for 
correct size of neutrino masses.
If we assume that all the charges are integer, then we have to take 
$l=-2$ to realize integer $t$. Under this assumption, the minimum value of
$n$ is 2 (namely $(\psi_1,\psi_2,\psi_3)=(5,4,2), t=3, h=-4)$, and we obtain 
essentially three solutions which satisfy the above
 necessary conditions:
($a'=3, c=-3,\bar c=0$),
($a'=3,4, c=-4,\bar c=-1$)
and ($a'=3,4, c=-5,\bar c=-2$). 
\footnote{
The last candidate is not so good because $c+\bar c=-7$ which may lead to 
smaller Cabbibo angle by a cancellation. }
We have some freedom to choose the charges 
$z, \bar z, h', c', \bar c', s$. Typical values are
$z=\bar z=-2, h'=5, c'=2-\bar c, \bar c'=2-c, s=-(c+\bar c)$.
\footnote{
If we adopt lopsided type neutrino mass matrix, the second condition
(\ref{t})
is replaced by
\begin{equation}
t=n+\frac{1}{2}(c-\bar c+4). \nonumber
\end{equation}
The minimum value of $n$ is also 2(namely,
 $(\psi_1,\psi_2,\psi_3)=(5,4,2), t=2, h=-4)$, and we have only one solution
$a'=3,4, c=-4,\bar c=0$, and
typical values of charges $z, \bar z, h', c', \bar c', s$ are
$z=\bar z=-2, h'=5, c'=2-\bar c, \bar c'=2-c, s=-(c+\bar c)$.
}

If we allow to take half integer charges,
\footnote{To adopt half integer charges with the FN field's charge
$\theta=-1$ becomes essentially the same as to adopt only integer 
charges with $\theta=-2$. If we have no singlet field with charge
$-1$, the model has naturally half integer charges with $U(1)_A$
normalization $\theta=-1$.}
 then the minimum value of $n$
satisfying the above necessary conditions becomes 3/2 (namely 
$(\psi_1,\psi_2,\psi_3)=(9/2,7/2,3/2), t=2, h=-3)$.
We can get only a solution:
$a'=3,c=-7/2,\bar c=1/2$. We have some freedom to choose the charges 
$z, \bar z, h', c', \bar c', s$. Typical values are
$z=\bar z=-2, h'=4, c'=3/2, \bar c'=11/2, s=3$.

\begin{figure}[htb]
\begin{center}
\leavevmode
\epsfxsize=110mm
\put(300,50){{\large $\bf{\log \mu (GeV)}$}}
\put(0,260){{\Large $\bf{\alpha^{-1}}$}}
\put(29,240){$\alpha_1^{-1}$}
\put(31,150){$\alpha_2^{-1}$}
\put(31,90){$\alpha_3^{-1}$}
\epsfbox{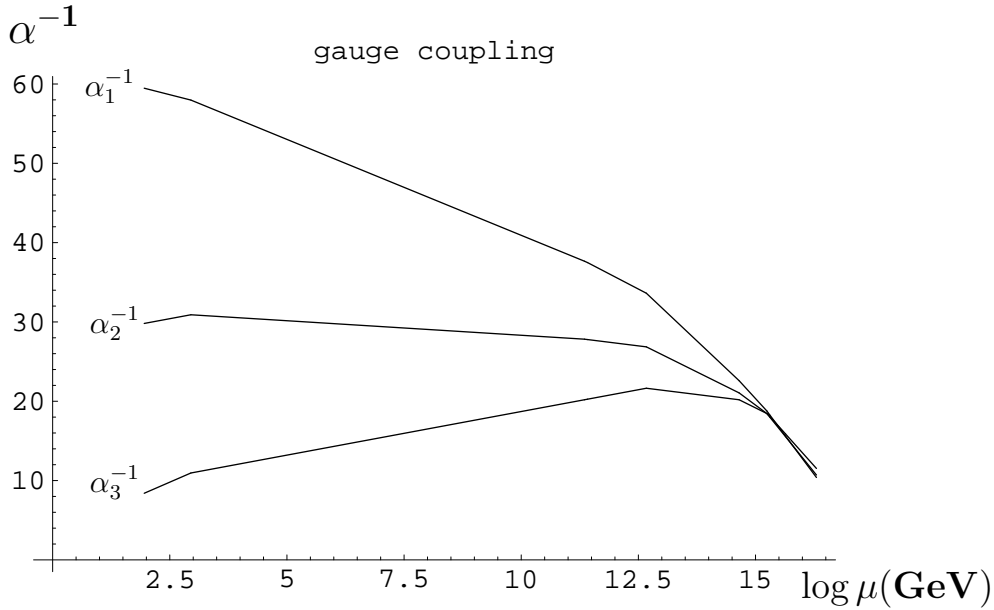}
\vspace{-2cm}
\caption{
Here we adopt $\alpha_1^{-1}(M_Z)=59.47$, $\alpha_2^{-1}(M_Z)=29.81$,
$\alpha_3^{-1}(M_Z)=8.40$, the SUSY breaking scale $m_{SB}\sim 1$ TeV and
the anomalous $U(1)_A$ charges: $a'=3$, $a=-1$, 
$(\psi_1,\psi_2,\psi_3)=(9/2,7/2,3/2)$, $t=2$, $h=-3$, $c=-7/2$,
$\bar c=1/2$, $z=\bar z=-2$, $h'=4$, $c'=3/2$, $\bar c'=11/2$ and $s=3$.
Using the ambiguities of coefficients $0.5\leq y\leq 2$,
three gauge couplings meet at around 
$\lambda^{-a}\Lambda_G\sim 5\times 10^{15}$ GeV. 
}
\label{fig_1}
\end{center}
\end{figure}

When  all the charges are determined, we can calculate the running
flows of gauge couplings (see Fig. \ref{fig_1}). 
Here we use the ambiguities of the coefficients $0.5\leq y\leq 2$.
It is shown that the three gauge couplings actually meet around
$\lambda^{-a}\Lambda_G\sim 5\times 10^{15}$ GeV.
\footnote{
This is not inconsistent with the discussion in ref. \cite{IKNY},
though they concluded that the coupling unification with a simple 
gauge group is generally impossible. The essential difference is
that we have not adopted their assumption $\Lambda\sim 10^{18}$ GeV.
}
Even the cases $n=2$, gauge coupling unification is possible, though
we have to use larger ambiguities of the coefficients.

In these cases, since the unification scale
$\Lambda_U\sim \lambda\Lambda_G$ becomes smaller than the usual GUT scale
$\Lambda_G\sim 2\times 10^{16}$ GeV, proton decay via dimension six operator
$p\rightarrow e^+\pi^0$
may be seen in near future. If we roughly estimate the lifetime of proton
using the formula in Ref.~\cite{hisano} and the recent result of 
the lattice calculation for the hadron matrix element parameter
$\alpha$\cite{lattice} 
\begin{equation}
\tau_p\sim 4.4\times 10^{34}\left(\frac{\Lambda_U}{10^{16}{\rm GeV}}\right)^4
\left(\frac{0.015}{\alpha}\right)^2  {\rm years},
\end{equation}
the lifetime of the proton in these cases becomes 
\begin{equation}
\tau_p\sim 2.8\times 10^{33} {\rm years}
\end{equation}
because the unification scale is around $5\times 10^{15}$ GeV.
It is interesting that this value of the lifetime is just around the 
present experimental limit
\cite{SKproton}
\begin{equation}
\tau_{exp}(p\rightarrow e^+\pi^0)>2.9\times 10^{33} {\rm year}.
\end{equation}
Of course, since we have an ambiguity of order one coefficients and of
the hadron matrix element parameter $\alpha$, and
the lifetime of proton is strongly dependent on the GUT scale and 
the parameter,
this prediction may not be so reliable. 
However, the above rough estimation gives us a strong motivation for
experiments of proton decay search, 
because the lifetime of nucleon via dimension
six operator
must be less than that in the usual SUSY GUT scenario. 

We have to comment on the proton decay via dimension five operators.
The effective colored Higgs mass is given by
$\lambda^{2h}\Lambda\sim 2\times 10^{20}$ GeV even if we take
$\Lambda=2\times 10^{16}$ GeV. Therefore the proton decay via dimension
five operator is still suppressed.

In $E_6$ unification case, the above analysis is a bit changed
as discussed in Ref.~\cite{maekawa}.
We have to introduce the effective anomalous $U(1)_A$ charges, which
are available only for the estimation of the mass determinants.
Actually in the above analysis, the charge of Higgs field must be
replaced by
\begin{equation}
h\rightarrow h+\frac{1}{4}(\phi-\bar \phi),
\end{equation}
where $\phi$ and $\bar \phi$ are the anomalous $U(1)_A$ charges of
$\Phi$ and $\bar \Phi$, whose VEVs 
$\VEV{\Phi}=\VEV{\bar \Phi}\sim \lambda^{-\frac{1}{2}(\phi+\bar \phi)}$
break $E_6$ into $SO(10)$. 
This is because the mass of the fields $D^c$ and $L$ is determined not
only by the charges but also by the VEV $\VEV{\Phi}$. \footnote{
Strictly speaking, even in $SO(10)$ unification case, we have to introduce
the effective charges for the mass matrices, because the mass term
between $({\bf \bar 5, 16})$ and $({\bf 5, 10})$ is dependent on the VEV
$\VEV{C}\sim \lambda^{-\frac{1}{2}(c+\bar c)}$. However, in the calculation
in this paper, these effects happen to be cancelled. If the Higgs doublet
$H_d$ originates from $({\bf \bar 5, 16})$, then these effects must be taken
into account. }
It is easily checked that all the effective charges can be defined 
consistently, though the effective charges for the mass matrices of different
representation are generically different even if they originate from the
same multiplet of $E_6$.
In principle, this modification can change the above situation of coupling
unification. Unfortunately the situation is not so improved but even worse, 
since 
$\phi-\bar \phi$ must be negative
for small $n$ in order to obtain the realistic quark and lepton mass 
matrices.
As discussed in Ref.~\cite{BM}, the conditions for obtaining the realistic
quark and lepton mass matrices are
\begin{equation}
c-\bar c=\phi-\bar\phi+1=2n-9-l,
\end{equation}
where we take $l=-1$ or $-2$ because $\Lambda\sim \Lambda_G$.
In $E_6$ DT splitting mechanism \cite{BMY},
Higgs $H$ is naturally unified into the multiplet $\Phi$,
namely $h=\phi=-2n$. 
In order to obtain the effective charge of Higgs 
$h+\frac{1}{4}(\phi-\bar \phi)=\frac{1}{4}(-6n-10-l)\sim 0$,
the small $n$ is required. 
From the condition $\phi+\bar \phi=-6n+10+l\leq -1$, the smallest value of 
$n$ becomes $3/2$ for $l=-2$. Then $\phi=-3$ and $\bar \phi=2$. 
In order to satisfy the economical condition for the $\mu$ problem
\begin{equation}
-1\leq 2h-(c+\bar c)+\frac{1}{2}(\phi-\bar \phi)\leq 1,
\label{mu}
\end{equation}
we adopt $c+\bar c=-8$, then $c=-6$ and $\bar c=-2$.
It is interesting that in this model we do not have to introduce 
R-parity because
half integer anomalous $U(1)_A$ charges can play the same role.
Unfortunately the effective Higgs mass becomes
$h+\frac{1}{4}(\phi-\bar \phi)=-\frac{17}{4}$, which is a bit larger than
the minimum value in $SO(10)$ unification case, though the gauge coupling
unification may be possible using larger ambiguities of order one coefficients.
Of course, we have to examine whether such a charge assignment is 
consistent with 
the DT splitting mechanism in $E_6$ unification or not,
that will be discussed in separate paper
\cite{BMY}.

\section{Discussions and Summary}
In this paper, we have examined the conditions for gauge coupling 
unification
with the anomalous $U(1)_A$ gauge theory and discussed several models which
satisfy the conditions. Since the unification scale and
the spectrum of super-heavy particles are determined only by the anomalous
$U(1)_A$ charges, the unification conditions are described by the charges.
We obtained a remarkable result that if all the fields except the MSSM fields 
have super-heavy masses, only a condition $h\sim 0$ realizes the gauge 
coupling unification. The unification 
scale becomes $\lambda^{-a}\Lambda_G$ and the cutoff scale becomes around
the usual GUT scale $\Lambda_G\sim 2\times 10^{16}$ GeV.
It is surprising that these results are independent on the details of the 
Higgs contents and their charge assignment. Therefore the predictions
are rather rigid, though we have some ambiguities of order 1 coefficients.

It is interesting that the unification scale is smaller than 
the usual GUT scale $\Lambda_G$, since $a<0$. Therefore, proton decay
through dimension six operator can be seen in future experiment.
Actually, if we adopt $a=-1$, the lifetime of nucleon
becomes around the present experimental limit.
Moreover, our scenario predicts 
smaller cutoff scale than the Planck scale.
One way to explain this discrepancy is to introduce extra dimension
in which only gravity modes can propagate. Such a structure has been examined
in the context of strongly coupled Heterotic string theory\cite{horava}.
It is interesting that the structure may give a solution for the
FCNC problem in SUSY breaking sector, if only gravity modes mediate the 
SUSY breaking effect from the  hidden brane to our visible brane.

\section{Acknowledgement}
We would like to thank M. Bando, T. Kugo, M. Yamaguchi and T. Yamashita for 
useful comments.


\begin{thebibliography}{99}
\bibitem{georgi}  H. Georgi and S.L. Glashow, {\it Phys. Rev. Lett.}
                  {\bf 32} (1974) 438.
\bibitem{SUSYGUT} E. Witten, {\it Nucl. Phys.} {\bf B188} (1981) 513;\\
                  S. Dimopoulos, S. Raby and F. Wilczek, {\it Phys. Rev.}
                  {\bf D24} (1981) 1681;\\
                  D. Dimopoulos and H. Georgi,{\it Nucl. Phys.} {\bf B193}
                  (1981) 150;\\
                  N. Sakai, {\it Z. Phys.} {\bf C11} (1981) 153.
\bibitem{SK}        Fukuda et al.(The Super-Kamiokande Collaboration),
                    {\it Phys. Lett.} {\bf B436} (1998) 33;
                    {\it Phys. Rev. Lett.} {\bf 81} (1998) 1562;
                    {\it Phys.Rev.Lett.} {\bf 86} (2001) 5656.
\bibitem{Sato}      J. Sato and T. Yanagida, {\it Phys.Lett.} {\bf B430}
                    (1998) 127.
\bibitem{Nomura}    Y. Nomura and T. Yanagida, {\it Phys.Rev.} {\bf D59}
                    (1999) 017303.
\bibitem{IKNY}      K.-I. Izawa, K. Kurosawa, Y. Nomura, T. Yanagida,
                    {\it Phys. Rev.} {\bf D60} (1999) 115016.
\bibitem{Bando}  M. Bando and T. Kugo, {\it Prog. Theor. Phys.} {\bf 101}
                 (1999) 1313; \\
                 M. Bando, T. Kugo and K. Yoshioka, {\it Prog. Theor. Phys.}
                 {\bf 104} (2000) 211.
\bibitem{Barr}   C.H. Albright, K.S. Babu and S.M. Barr,  {\it
Phys.Rev.Lett.}
                 {\bf 81} (1998) 1167;\\
                 C.H. Albright and S.M. Barr, {\it Phys.Rev.Lett.} {\bf 85}
                  (2000) 244;
                 {\it Phys.Rev.} {\bf D62} (2000) 093008;
                 {\it Phys.Lett.} {\bf B461} (1999) 218;
                 {\it Phys.Lett.} {\bf B452} (1999) 287;
                 {\it Phys.Rev.} {\bf D58} (1998) 013002.
\bibitem{Shafi} Q. Shafi and Z. Tavartkiladze, {\it Phys. Lett.} {\bf B487}
                 (2000) 145.
\bibitem{DTsplitting} E. Witten, {\it Phys. Lett.} {\bf B105} (1981) 267;\\
                      A. Masiero, D.V. Nanopoulos, K. Tamvakis and T.Yanagida,
                      {\it Phys. Lett.} {\bf 115} (1982) 380;\\
                      K. Inoue, A. Kakuto and T. Takano,
                      {\it Prog. Theor. Phys.} {\bf 75} (1986) 664;\\
                      E. Witten, {\it Nucl. Phys.} {\bf B258} (1985) 75;\\
                      T. Yanagida, {\it Phys. Lett.} {\bf B344}  (1995) 211;\\
                      Y. Kawamura, {\it Prog. Theor. Phys.} {\bf 105} (2001)
                      691;{\it ibid} {\bf 105} (2001) 999.
\bibitem{DW}       S. Dimopoulos and F. Wilczek, NSF-ITP-82-07;\\
                   M. Srednicki, {\it Nucl. Phys.} {\bf B202} (1982) 327.
\bibitem{BarrRaby} S.M. Barr and S. Raby, {\it Phys. Rev. Lett.}
                   {\bf 79} (1997) 4748.
\bibitem{Chako}    Z. Chacko and R.N. Mohapatra,
                   {\it Phys.Rev.} {\bf D59} (1999) 011702;
                   {\it Phys.Rev.Lett.} {\bf 82} (1999) 2836.
\bibitem{complicate} K.S. Babu and S.M. Barr, {\it Phys. Rev} {\bf D48}
                    (1993) 5354;{\it ibid} {\bf D50} (1994) 3529.
\bibitem{maekawa} N. Maekawa, {\it Prog. Theor. Phys.} {\bf 106} (2001) 401
                  (hep-ph/0104200).
\bibitem{BM}      M. Bando and N. Maekawa, hep-ph/0109018 to appear in 
                  {\it Prog. Theor. Phys.} 
\bibitem{U(1)}    E.~Witten, {\it Phys. Lett.} {\bf B149} (1984),351;\\
                  M.~Dine, N.~Seiberg and E.~Witten,
                  {\it Nucl. Phys.} {\bf B289} (1987), 589;\\
                  J.J.~Atick, L.J.~Dixon and A.~Sen,
                  {\it Nucl. Phys.} {\bf B292} (1987),109;\\
                  M.~Dine, I.~Ichinose and N.~Seiberg,
                  {\it Nucl. Phys. } {\bf B293} (1987),253.
\bibitem{GS}      M.~Green and J.~Schwarz,
                  {\it Phys. Lett.} {\bf B149} (1984),117.
\bibitem{Ibanez}  L.~Ib\'a\~nez and G.G.~Ross,
                  {\it Phys. Lett.} {\bf B332} (1994),100;\\
                  P.~Bin\'etruy and P.~Ramond,
                  {\it Phys. Lett. } {\bf B350} (1995),49;\\
                  E.~Dudas, S.~Pokorski and C.A.~Savoy,
                  {\it Phys. Lett.} {\bf B356} (1995),45;\\
                  P.~Bin\'etruy, S.~Lavignac and P.~Ramond,
                  {\it Nucl. Phys. } {\bf B477} (1996),353.
\bibitem{Ramond}  P.~Bin\'etruy, S.~Lavignac, S.~Petcov and P.~Ramond,
                  {\it Nucl. Phys. } {\bf B496} (1997),3.
\bibitem{Dreiner} H.~Dreiner, G.K.~Leontaris, S.~Lola, G.G.~Ross and 
                  C.~Scheich,{\it Nucl. Phys. } {\bf B436} (1995),461.
\bibitem{maekawa2} N. Maekawa, {\it Phys. Lett.} {\bf B521} (2001) 42 
                  (hep-ph/0107313).
\bibitem{FN}      C.D. Froggatt and H.B. Nielsen,
                  {\it Nucl. Phys.} {\bf B147} (1979) 277.
\bibitem{CKM}     M. Kobayashi and T. Maskawa, {\it Prog. Theor. Phys.}
                  {\bf 49} (1973) 652.
\bibitem{MNS}     Z. Maki, M. Nakagawa and S. Sakata,
                  {\it Prog. Theor. Phys.} {\bf 28} (1962) 870.
\bibitem{CHOOZ}   The CHOOZ Collaboration, M. Appollonio et al.,
                  {\it Phys. Lett.} {\bf B420}(1998) 397.
\bibitem{MSW}     L. Wolfenstein, {\it Phys. Rev.} {\bf D17} (1978) 2369;\\
                  S.P. Mikheev and A. Smirnov, {\it Yad. Fiz.} {\bf 42}
                  (1985) 1441; {\it Nuovo Cim.} {\bf 9C} (1986) 17.
\bibitem{Valle}   J.W.F. Valle, astro-ph/0104085; \\
                  M.C. Gonzalez-Garcia, M. Maltoni, C. Pena-Garay and
                  J.W.F. Valle, {\it Phys.Rev.} {\bf D63} (2001) 033005.
\bibitem{lopsided} F. Vissani, {\it JHEP} {\bf 9811} (1998) 025;
                   {\it Phys. Lett.} {\bf B508} (2001) 79. \\
                   J. Sato and T. Yanagida, {\it Phys. Lett.} {\bf B493}
                   (2000) 356; \\
                   M. Tanimoto, {\it Phys. Lett.} {\bf B501} (2001) 231.
\bibitem{yamaguchi} S. Komine and M. Yamaguchi, hep-ph/0110032.
\bibitem{kawamura} M. Drees, {\it Phys. Lett.} {\bf B181} (1986) 279;\\
                   Y. Kawamura, H. Murayama and M. Yamaguchi,
                 {\it Phys.Lett.} {\bf B324} (1994) 52; 
                  {\it Phys.Rev.} {\bf D51} (1995) 1337.
\bibitem{hisano}  J. Hisano, H. Murayama and T. Yanagida, {\it Nucl. Phys.}
                  {\bf B402} (1993) 46.
\bibitem{lattice} JLQCD Collaboration, S. Aoki et al., {\it Phys. Rev.}
                  {\bf D62} (2000) 014506.
\bibitem{SKproton} Super-Kamiokande Collaboration, {\it Phys. Rev. Lett.}
                  {\bf 81} (1998) 3319; {\it ibid} {\bf 83} (1999) 1529.
\bibitem{BMY}      M. Bando, N. Maekawa and T. Yamashita, in preparation.
\bibitem{horava}  P. Horava and E. Witten, {\it Nucl. Phys.} {\bf 460}
                  (1996) 506; {\it ibid} {\bf 475} (1996) 94.
\end{thebibliography}
\end{document}